\documentclass[prb,preprint]{revtex4-1} 

\usepackage{graphicx}
\usepackage{dcolumn}
\usepackage{bm}

\usepackage{fancyhdr}

\usepackage{amsmath}
\usepackage{amsthm}
\usepackage{amsfonts}
\usepackage{amssymb}
\usepackage{latexsym}
\usepackage{mathrsfs}
\usepackage{tikz}
\usepackage[mathscr]{eucal}
\usepackage{lipsum}

\usepackage{calligra}
\DeclareMathAlphabet{\mathcalligra}{T1}{calligra}{m}{n}
\DeclareFontShape{T1}{calligra}{m}{n}{<->s*[2.2]callig15}{}
 
\newcommand{\scriptr}[1]{\ensuremath{\mathcalligra{#1}}}

\usepackage{enumitem}
\setlist{  
  listparindent=\parindent,
  parsep=0pt,
}

\begin{document}


\title{An elementary argument for the magnetic field outside a solenoid}

\author{Aritro Pathak}
\affiliation{%
 Department of Mathematics \\ Brandeis University
}%



\date{\today}

\pacs{Valid PACS appear here}

\begin{abstract}
    
    The evaluation of the magnetic field inside and outside a uniform current density infinite solenoid of uniform cross-section is an elementary problem in classical electrodynamics that all undergraduate Physics students study.  Symmetry properties of the cylinder and the judicious use of Ampere's circuital law leads to correct results; however it does not explain why the field is non zero for a finite length solenoid, and why it vanishes as the solenoid becomes infinitely long. An argument is provided in (American Journal of Physics 69, 751 (2001)) by Farley and Price, explaining how the magnetic field behaves outside the solenoid and not too far from it, as a function of the length of the solenoid. A calculation is also outlined for obtaining the field just outside the circular cross section solenoid, in the classic text Classical Electrodynamics by J.D.Jackson, 3rd ed, (John Wiley and Sons, INC)  Problems 5.3, 5.4, 5.5. The purpose of this letter is to provide an elementary argument for why the field becomes negligible as the length of the solenoid is increased.  An elementary calculation is provided for the field outside the solenoid, at radial distances large compared to the linear dimension of the solenoid cross section.

\end{abstract}

\maketitle

\section{Introduction}

Consider a solenoid of length $L$ and arbitrary cross section of area $A$. Every high school and beginning undergraduate student is taught that an infinitely long ideal solenoid produces uniform magnetic field within the core of the solenoid and that the field outside the solenoid vanishes. The exact field outside the infinite solenoid has been extensively studied in literature\cite{Dasgupta}. The exact solution for the finite solenoid has also been studied, notably in an article from 1960.\cite{Nasa} Informally, the two ends of the long solenoid behave like two magnetic poles, each of which produce fields that fall off with the inverse square of distance, and thus the fields outside are small when the length of the solenoid is increased, keeping the current fixed. Typical introductory texts use symmetry arguments to prove at first that the fields have to be parallel to the axis, and then use Ampere's Law to find the magnetic field outside and inside the solenoid.\cite{Halliday}\cite{Walter}. 

This standard textbook treatment, while correct, does not explain how the fields due to each individual coil of the solenoid add up to produce no net field at any point outside the solenoid, in the limit of the solenoid length going to infinity. There are some introductory texts that try to address this by saying the density of field lines outside the solenoid decreases as the length of the solenoid is increased \cite{Wolfson}, which is not based on any quantitative analysis. Farley and Price\cite{Farley} argue that the magnetic field just outside the finite solenoid is roughly constant in the central region; and only then they are able to use the reciprocity theorem of mutual inductance in an elegant way, to estimate the magnetic field outside the finite solenoid, on the median plane. 

\begin{figure}
    \centering
    \includegraphics[scale=0.45]{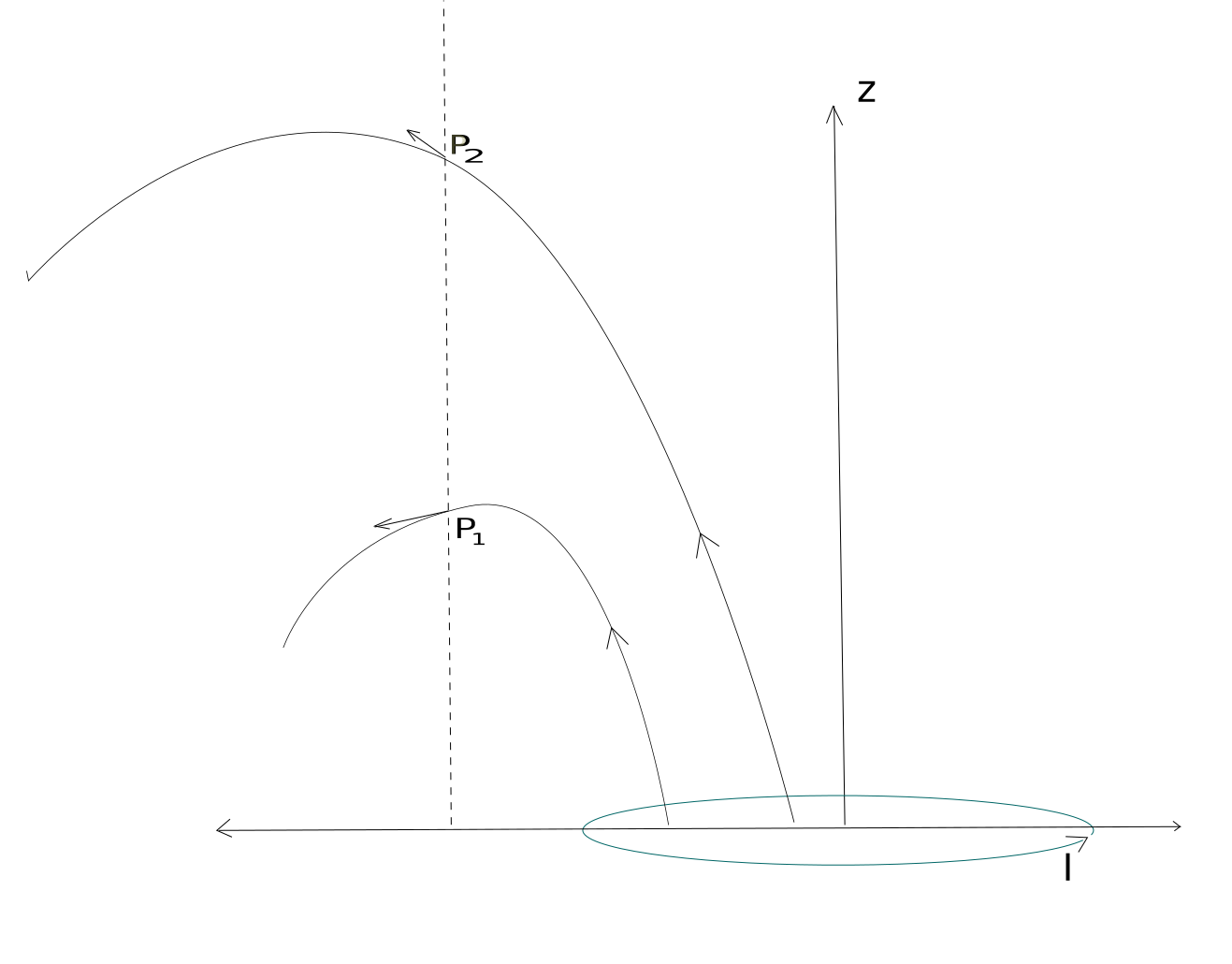}
    \caption{The magnetic field due to a single coil carrying current I, at the two points $P_{1}, P_{2}$.}
\end{figure}

We first provide an elementary heuristic argument in Section II, explaining why the field anywhere outside the solenoid has to become vanishingly small, as the length of the solenoid is increased. The key point to note is, the field at any point $P$ outside the solenoid due to coils closer to it , and the field at the same point due to coils farther away from it, have opposite direction along the axis, as illustrated in Fig 2. To corroborate the arguments of Section II, we show a calculation for the magnetic field outside the solenoid, at radial distances $\rho>>\sqrt{A}$, in Section III. Our calculation directly shows that the magnetic field strength at distances much greater than the solenoid radial dimension is roughly constant, and we do not need to a priori argue it before being able to calculate it quantitatively, as in the case of the calculation in Ref.\cite{Farley}. It is indeed surprising that our simple and easily motivated analysis does not appear  anywhere in the literature.

One caveat for the usual textbook treatment is that for any actual solenoid, there has to be a component of current flowing along the surface of the solenoid, in the axial direction. Because of this, there is a feeble magnetic field in the azimuthal direction surrounding the solenoid.\cite{Serway},\cite{Minoru} For the purposes of this letter, we neglect this tiny field; we treat the solenoid as a stack of tightly packed uniform cross-section identical coils carrying the same current.

\section{A heuristic approach}

\begin{figure}
    \centering
    \includegraphics[scale=0.22]{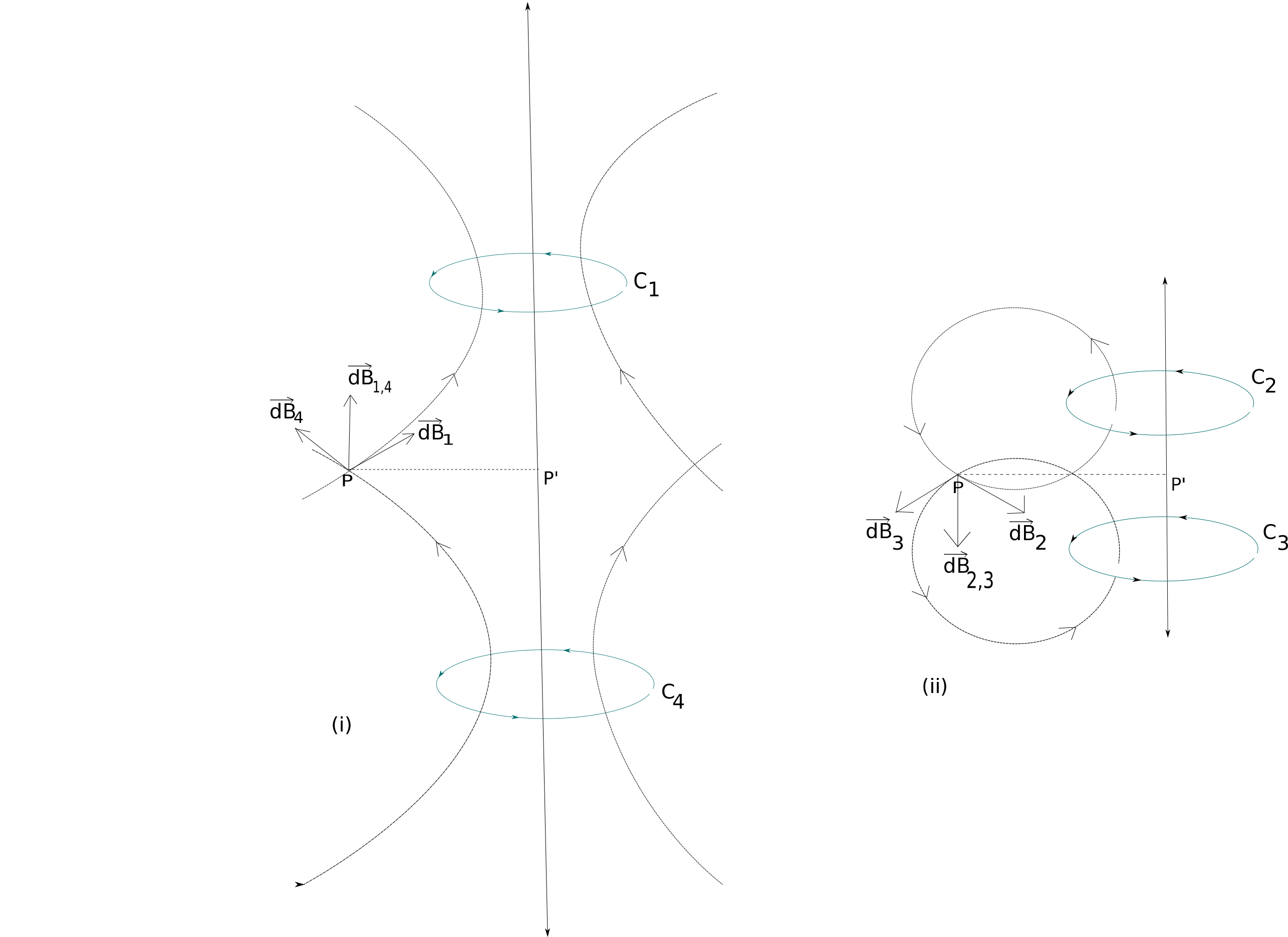}
    \caption{ FIG 2. The Magnetic field at point $ P$ : (i) due to coils farther away from $P$ on the left; and (ii) due to coils closer to $P$ on the right.\cite{coment}}
    \label{fig:my_labell}
\end{figure}

The field lines due to an individual circular current carrying coil are shown in Fig 1. Both points $P_{1}$ and $P_{2}$ are at the same radial distance from the axis of the coil, with $P_{2}$ being vertically farther away from the coil than $P_{1}$. The z-component of the magnetic field at the point $P_{1}$ is negative, while that at the point $P_{2}$ is positive. 
    
We draw a circular coil for simplicity, but our arguments hold for any arbitrary solenoid cross section.

Consider any arbitrary point P outside the solenoid and left of it, where we are to find the magnetic field. This point might be arbitrarily close to, or very far away from the solenoid; the subsequent argument remains unchanged.The current in each of the coils of the solenoid flows counter clockwise when viewed from the top. 

The perpendicular to the axis, passing through the point P, touches the axis at the point P', as shown in Fig 2. The line PP' and the axis of the solenoid determine the plane under consideration. Consider first the individual loops of the solenoid far from the point P'. As shown in Figure 2(i), a loop far below P'(the loop $C_{4}$  in Fig. 2) produces the field $\vec{dB}_{4}$ at P. Similarly, a symmetrically placed loop far above P' (the loop $C_{1}$ in Fig. 2) produces the field $\vec{dB}_{1}$ at $P$. The resultant of these two fields is $\vec{dB}_{1,4}$ that points up along the axis, i.e. along the direction of flow of current. The radial component of this combined field at $P$ cancels out due to symmetry. The distance of these coils from the point P' is large, and their individual contributions fall of as the inverse cube of this distance; but the number of loops producing this field up along the axis extends to infinity for an infinite solenoid (both above and below P'), and so we have a significant contribution.


We are left to deal with the loops close to $P'$. The field lines from these loops, that touch upon $P$, produce the fields $\vec{dB}_{2}$(due to the coil $C_{2}$) and $\vec{dB}_{3}$(due to the coil $C_{3}$), so that the resultant is the field $\vec{dB}_{1,2}$. The contributions due to each of the two loops $C_{1}, C_{2}$ is fairly large, while the number of such loops producing the fields in the downward direction are fairly small.

In the regime where $\rho>>\sqrt{A}$, we show in the next section that we get the downward field at $P$, only for loops till a distance $d\leq \rho/\sqrt{2}$ on either side of P' along the axis. For a finite solenoid of small length, we only have significant contributions from the loops producing the downward field. There won’t be enough loops far away from the point P that produces the upward field, because the solenoid is finite. For the infinite solenoid, we end up having a negligible field because these two effects counteract each other well.

 If our observation point is far away, the number of loops producing the downward field increases while the number of loops producing the upward field still extends to infinity for the infinite solenoid. The net effect remains the same.

\section{Calculation of the field in the region $\rho>>\sqrt{A}$}

For the finite solenoid, the two length scales of the problem are the radial dimension of the cross section $\sqrt{A}$ and the length $L$ of the solenoid. At distances much larger than $L$, the entire solenoid itself is treated as a point magnetic dipole, as noted in Ref \cite{Farley}. In our case, $L>>\sqrt{A}$. Even at radial distances comparable to $L$ but much larger than $\sqrt{A}$, we can treat each individual current loop as a point dipole, and add up the field due to each current loop. The essential feature of the argument of the previous section shows up in the calculation; the  axial component of the field due to a particular current loop switches sign after we move sufficiently far away from it along the z-direction. The analysis in this section is similar to that required for the configuration presented in Problem 5.61 in Griffiths's Introduction to Electrodynamics.\cite{Griffiths}

The solenoid has $n$ turns per unit length. For a loop of infinitesimal thickness $dz'$ along the axis, there is a current $nI dz'$ through this loop.The magnetic moment of this loop is $\vec{m}=nIAdz'\hat{z}$. The magnetic field of this point dipole is:

$$\vec{dB}(\vec{\scriptr{r}})=\frac{\mu_{0}}{4\pi}\frac{3(\vec{m}\cdot \hat{\scriptr{r}})\hat{\scriptr{r}}-\vec{m}}{\scriptr{r}^{3}} \ \ \ \ \ \ \ \ (1)       $$

Here $\vec{\scriptr{r}}=\vec{r}-\vec{r'}=(z-z')\hat{z}+\rho\hat{\rho}$ is the vector to the point $P$ from the coil element, as shown in Fig. 3. The solenoid lies along the z-axis, and stretches from $-L/2$ to $L/2$. 

The radial component of the field at the point $P$ due to the coil element at $z'$ is:

$$ \vec{dB}_{\rho}(\vec{r})=\frac{3\mu_{0}nIA\rho dz'}{4 \pi}\frac{(z-z')}{\scriptr{r}^{5}}\hat{\rho} \ \ \ \ \ \ \  \ (2)$$

The radial component of the field vanishes for an infinite solenoid due to symmetry, as can be seen by integrating the above equation for all values of $z'$ from $-\infty$ to $\infty$. Even for the finite solenoid, this component is small, as can be argued from the methods of Ref \cite{Farley}. We henceforth ignore this radial component.

The axial field at point $P$ due to the coil element at $z'$ is given by:

$$ \vec{dB}_{z}(\vec{r})=  \frac{nIA\mu_{0}dz'}{4\pi} (\frac{2(z-z')^{2}-\rho^{2}}{\scriptr{r}^{5}})\hat{z} \ \ \ \ \ \ \ \ (3)$$

\

Whenever $|z-z'|/\rho<1/\sqrt{2}$ the field points in the negative $z$ direction, and switches sign for a greater vertical separation, as argued in Section II.

\begin{figure}
\centering

\begin{tikzpicture}

\draw[black, thick,->] (0,5.0)--(0,6.2) node[above] (TextNode) {$\hat{z}$};
\draw[black, thick,->] (0,-2.0)--(0,-3.0);
\draw[black, thick,->] (0.0,0)--(7.0,0) node[above] (TextNode) {$\hat{\rho}$};
\draw[black, thick] (0.1,-2.0) -- (0.1,5.0) ;
\draw[black, thick] (-0.1,-2.0) -- (-0.1,5.0) ;
\draw[black, thick] (-0.1,5.0) -- (0.1,5.0) ;
\draw[black, thick] (-0.1,4.67) -- (0.1,4.67) ;
\draw[black, thick] (-0.1,4.33) -- (0.1,4.33) ;
\draw[black, thick] (-0.1,4.0) -- (0.1,4.0) ;
\draw[black, thick] (-0.1,3.67) -- (0.1,3.67) ;
\draw[black, thick] (-0.1,3.33) -- (0.1,3.33) ;
\draw[black, thick] (-0.1,3.0) -- (0.1,3.0) ;
\draw[black, thick] (-0.1,2.67) -- (0.1,2.67) ;
\draw[black, thick] (-0.1,2.33) -- (0.1,2.33) ;
\draw[black, thick] (-0.1,2.0) -- (0.1,2.0) ;
\draw[black, thick] (-0.1,1.67) -- (0.1,1.67) ;
\draw[black, thick] (-0.1,1.33) -- (0.1,1.33) ;
\draw[black, thick] (-0.1,1.0) -- (0.1,1.0) ;
\draw[black, thick] (-0.1,0.67) -- (0.1,0.67) ;
\draw[black, thick] (-0.1,0.33) -- (0.1,0.33) ;
\draw[black, thick] (-0.1,0.0) -- (0.1,0.0) ;
\draw[black, thick] (-0.1,-0.33) -- (0.1,-0.33) ;
\draw[black, thick] (-0.1,-0.67) -- (0.1,-0.67) ;
\draw[black, thick] (-0.1,-1.0) -- (0.1,-1.0) ;
\draw[black, thick] (-0.1,-1.33) -- (0.1,-1.33) ;
\draw[black, thick] (-0.1,-1.67) -- (0.1,-1.67) ;

\draw[black, thick] (-0.1,-2.0) -- (0.1,-2.0) ;
\fill[blue!25!,opacity=.3] (-0.1,-2) rectangle (0.1,5) ;
\draw[black, thick, ->] (0.0,1.2) -- (5.0,4.0) node[ midway, right] (TextNode) {$ \LARGE{\vec{\scriptr{r}}}$};
\draw[black, thick, ->] (0,0) -- (5,4) node[ midway, right] (TextNode) {$ \ \ \vec{r}$};
\draw[black, thick, ->] (0,0) -- (0,1.2) node[ midway, left] (TextNode) {$ \ \ \vec{r'}$};
\draw[black, dashed] (0,4) -- (5,4) ;
\filldraw[black] (5,4) circle (2pt)
node[anchor=west] {$P$};
\draw[black, dashed,<->] (0,-0.5) -- (5,-0.5) node[midway,below] (TextNode) {$\rho$};

\filldraw[black] (0,4) circle (2pt)
node[anchor=east] {$P'$};

\draw[black, dashed, <->] (5.0,0) -- (5.0,4.0) node [midway, right] (TextNode) {$z$};
\draw[black, dashed, <->] (-0.8,0) -- (-0.8,1.2) node [midway, left] (TextNode) {$z'$};

\end{tikzpicture}

\caption{FIG 3: The coordinates used in Section III are shown. The solenoid extends in both directions along the vertical $z$ axis. The magnetic field is evaluated at point P.}

\end{figure}
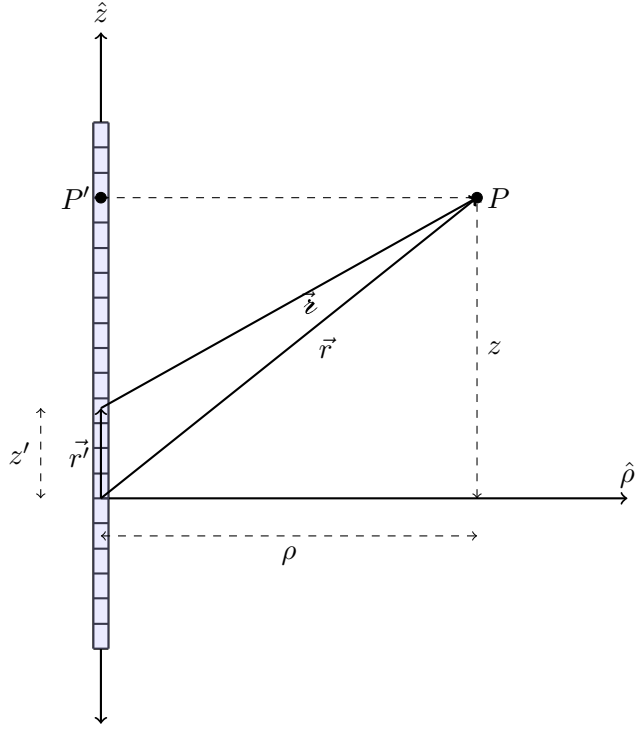

To find the net field at $P$, we integrate the previous expression for magnetic field between the limits $z'=L/2$ and $z'=-L/2$. This gives us the integral:

$$B_{z}(\vec{r})= - \frac{nIA\mu_{0}}{4\pi}\int_{z+L/2}^{z-L/2}(\frac{2t^{2}-\rho^{2}}{(t^{2}+\rho^{2})^{5/2}}) dt \ \ \ \ \ \ \ \ (4)$$

This integral is exactly solved with the substitution $t=\rho \ \tan \theta$. The antiderivative of the integrand in Equation.3 is $-t/(t^{2}+\rho^{2})^{3/2}$.  We thus get:

$$B_{z}(\vec{r})=  \frac{nIA\mu_{0}}{4\pi}\Big[\frac{z-L/2}{((z-L/2)^{2}+\rho^{2})^{3/2}}-\frac{z+L/2}{((z+L/2)^{2}+\rho^{2})^{3/2}}\Big] \  \ \   (5)$$

It is clear that in the limit $L\rightarrow \infty$, this expression becomes 0, i.e the field at $P$ is 0 for the infinite solenoid. 

For the finite solenoid, for points $P$ near the central region so that $|z|<<L/2$
and radial distance $\rho<<L$, we can compare this result with the expression of Ref \cite{Farley}. 

With the given assumptions, our expression becomes, to lowest orders in $\rho^{2}/(\frac{L}{2}+z)^{2}$ and $\rho^{2}/(\frac{L}{2}-z)^{2}$:

$$B_{z}(\vec{r})\approx  \frac{nIA\mu_{0}}{4\pi}\Big[\frac{(z-L/2)}{(L/2-z)^{3}}(1-\frac{3}{2}\frac{\rho^{2}}{(z-L/2)^{2}})-\frac{(z+L/2)}{(z+L/2)^{3}}(1-\frac{3}{2}\frac{\rho^{2}}{(z+L/2)^{2}})\Big] \ \ \ (5)$$

The terms containing $\rho^{2}/(\frac{L}{2}+z)^{4}$ and $\rho^{2}/(\frac{L}{2}-z)^{4}$ have a negligible contribution, and thus there is no non- negligible dependence of the field on radial distance\cite{Farley}. We get,

$$B_{z}(\vec{r})\approx -\frac{nIA\mu_{0}}{4\pi}(\frac{1}{(L/2 -z)^{2}}+\frac{1}{(L/2+z)^{2}}) \ \ \ \  \ (6)$$

Further, to order $(z/L)^{2}$:

$$B_{z}(\vec{r})\approx -\frac{2\mu_{0}nIA}{\pi L^{2}}(1+12(\frac{z}{L})^{2}+ O((z/L)^{4})) \ \ \ \ \ (7)$$

This gives the result for distances $L>>\rho>>\sqrt{A}$ and $|z|<<L/2$. Farley and Price's calculation \cite{Farley} is done only in the plane $z=0$, and only the first constant term of Eq. 7 was obtained.\cite{comment}

\section{Conclusion}

We have furnished a simple argument explaining how the magnetic field behaves outside a uniform current density solenoid, and provided an elementary calculation underpining the argument.


\begin{thebibliography}{0}%
\makeatletter
\providecommand \@ifxundefined [1]{%
 \@ifx{#1\undefined}
}%
\providecommand \@ifnum [1]{%
 \ifnum #1\expandafter \@firstoftwo
 \else \expandafter \@secondoftwo
 \fi
}%
\providecommand \@ifx [1]{%
 \ifx #1\expandafter \@firstoftwo
 \else \expandafter \@secondoftwo
 \fi
}%
\providecommand \natexlab [1]{#1}%
\providecommand \enquote  [1]{``#1''}%
\providecommand \bibnamefont  [1]{#1}%
\providecommand \bibfnamefont [1]{#1}%
\providecommand \citenamefont [1]{#1}%
\providecommand \href@noop [0]{\@secondoftwo}%
\providecommand \href [0]{\begingroup \@sanitize@url \@href}%
\providecommand \@href[1]{\@@startlink{#1}\@@href}%
\providecommand \@@href[1]{\endgroup#1\@@endlink}%
\providecommand \@sanitize@url [0]{\catcode `\\12\catcode `\$12\catcode
  `\&12\catcode `\#12\catcode `\^12\catcode `\_12\catcode `\%12\relax}%
\providecommand \@@startlink[1]{}%
\providecommand \@@endlink[0]{}%
\providecommand \url  [0]{\begingroup\@sanitize@url \@url }%
\providecommand \@url [1]{\endgroup\@href {#1}{\urlprefix }}%
\providecommand \urlprefix  [0]{URL }%
\providecommand \Eprint [0]{\href }%
\providecommand \doibase [0]{http://dx.doi.org/}%
\providecommand \selectlanguage [0]{\@gobble}%
\providecommand \bibinfo  [0]{\@secondoftwo}%
\providecommand \bibfield  [0]{\@secondoftwo}%
\providecommand \translation [1]{[#1]}%
\providecommand \BibitemOpen [0]{}%
\providecommand \bibitemStop [0]{}%
\providecommand \bibitemNoStop [0]{.\EOS\space}%
\providecommand \EOS [0]{\spacefactor3000\relax}%
\providecommand \BibitemShut  [1]{\csname bibitem#1\endcsname}%
\let\auto@bib@innerbib\@empty
\end{thebibliography}%


\begin{thebibliography}{99}

\bibitem{Farley} Jason Farley, Richard H. Price; \textit{Field just outside a long solenoid}  American Journal of Physics 69, 751 (2001).

\bibitem{Dasgupta} Basab Dasgupta, \textit{Magnetic field due to a solenoid} American Journal of Physics 52, 258 (1984) ; \ Keith Fillmore,\textit{Magnetic field of a noncircular solenoid}  American Journal of Physics 53, 782 (1985); \ Olivier Espinosa,  Viktor Slusarenko, \textit{The magnetic field of an infinite solenoid} American Journal of Physics 71, 953 (2003).

\bibitem{Nasa} Edmund E. Callaghan and Stephen H. Maslen, \textit{The Magnetic Field of a Finite Solenoid} NASA Technical Note D-465 (1960).


\bibitem{Halliday} Halliday Resnick Krane, \textit{Physics, Volume 2} (Fifth Edition) pp.758-760, 762-763.

\bibitem{Walter} Walter Hauser, \textit{Note on ‘‘Magnetic field due to a solenoid’’}  American Journal of Physics 53, 774 (1985).

\bibitem{Wolfson} R. Wolfson and J. M. Pasachoff, Physics, 3rd ed. ��Addison–Wesley, Reading, MA, 1999, Fig 30-28.

\bibitem{Serway} Raymond A. Serway, John W. Jewett, Jr. \textit{Physics for Scientists and Engineers with Modern Physics} 9th ed, (Brooks/Cole Cengage Learning) pp. 915-916.

\bibitem{Minoru} Minoru Harada, \textit{On the magnetic field of a solenoid}  American Journal of Physics 54, 1065 (1986).

\bibitem{coment} For the sake of argument, the field lines due to individual current loops are shown in either diagram; it should be clear the actual field lines for a configuration of multiple current loops do not resemble what is drawn in this figure.

\bibitem{Griffiths} David J. Griffiths, Introduction to Electrodynamics , 3rd ed. (Pearson, Upper Saddle River, NJ,2013) pp.254.

\bibitem{comment} Farley and Price's approach also works for points off the median plane, but close to it. There the assumption of axial field works well, and we can get exactly the expression in Eq. 7, when again the a priori assumption is made that the field is approximately constant outside the solenoid in our domain.


 
 




\end{thebibliography}
\end{document}